\documentstyle[twocolumn,aps,prl]{revtex}
\begin{document}

\draft
\title
{\bf Transient tunneling current of single electron transistors}

\author{ David M.-T. Kuo, Pei-Wen Li and W. T. Lai}
\address{Department of Electrical Engineering, National Central
University, Chung-Li, Taiwan, 320, Republic of China}

\date{\today}
\maketitle

\begin{abstract}
The transient tunneling current of single electron transistors
(SETs) is theoretically investigated. The time-dependent current
formula given by Jauho, Wingreen and Meir [Phys. Rev. B 50, 5528
(1994)] is applied to study the temperature effect on the
transient current through a single quantum dot embedded into
asymmetry barrier. It is found that the tunneling rate ratio
significantly influences the feature of transient current.
Finally, the oscillation structures on the exponential growth
transient current of single hole transistors composed of germanium
quantum dots is analyzed.
\end{abstract}

\section{Introduction}
The transport properties of single electron transistors (SETs)
have been extensively studied theoretically and
experimentally.$^{1-6}$ The main structure of SETs consists of a
single quantum dot (QD) and three electrodes (source, drain and
gate). The manipulation of SETs is based on Coulomb blockade
effect arising from the particle interactions of QD. From the
practical point of view, it is important for SETs to operate at
room temperature. Therefore, the size of QD is required to be less
than $10~nm$ for silicon (Si) or germanium (Ge) semiconductor QDs.
In such size range of QD, the quantum confinement effects and
charging energies of QDs are larger than the thermal energy of
room temperature. Recently, the Coulomb oscillation and staircase
features of tunneling current of room temperature SETs have been
reported by several groups, where Si or Ge QDs are embedded into
$SiO_2$ matrix$^{4-6}$.

Even though it is difficult to align a single QD of nanometer with
electrodes in the fabrication of individual SETs, several methods
were used to solve this difficulty. Nevertheless, those methods
still can not precisely control the barrier thickness, which
significantly influences the tunneling time for electrons to
access QDs. Further understanding for the location of QDs, the
implementation technology of SETs using thermal oxidation
method$^{6}$ can be improved. The measurement of transient current
can provide above information. In this study we apply the current
formula of Ref. [2] to examine the transient current of the SET
with asymmetrical tunneling rates at room temperature. It is found
that the transient current exhibits exponential decay when
electrons are injected into the QDs from the left electrode and
$\Gamma_L/\Gamma_R < 1$, where $\Gamma_{L/R}$ represents the
tunneling rate for electrons from the left (right) electrode to
the QD, on the other hand the transient current exhibits the
exponential growth as $\Gamma_L/\Gamma_R > 1$. In addition to
theoretical studies of transient current, we report the
experimental measurement of transient current of single hole
transistor (SHT).

\section{Formalism}
The transient current of a SET consisted of a single QD and three
electrodes was theoretically derived by Jauho, Wingreen and
Meir.$^{2}$ When applied voltage is insufficient to overcome the
charging energies arising from electron-electron repulsion
interactions in the QD,the time-dependent tunneling current
through the ground state of QD can be expressed as (ref.[2])
\begin{equation}
J(t)=\frac{1}{2}[J^{in}_L(t)+J^{out}_R(t)-J^{out}_{L}(t)-J^{in}_R
(t)],
\end{equation}
where
\begin{equation}
J^{out}_{L/R}(t)=\frac{-e}{\hbar} \Gamma_{L/R} N(t)
\end{equation}
and
\begin{equation}
J^{in}_{L/R}(t)=\frac{-e}{\hbar} \Gamma_{L/R} \int
\frac{d\epsilon}{\pi}  f_{L/R}(\epsilon) Im A_{C}(\epsilon,t).
\end{equation}
According to Eq. (1), there are four components for the net
current from the left electrode to the right electrode. The
current of Eq. (2), $J^{out}_{L/R}(t)$, denotes the current
flowing out from the QD to the left (right) electrode. This
current results from the time-dependent electron occupation number
of QD, which is given by
\begin{equation}
N(t)=\int \frac{d\epsilon}{2\pi} (\Gamma_L f_L(\epsilon)+\Gamma_R
f_R(\epsilon)) |A_C(\epsilon,t)|^2.
\end{equation}
$J^{in}_{L/R}(t)$ denotes the current flowing into the QD from the
left (right) electrode. Obviously, $J^{in}_{L/R}(t)$ is determined
by the time-dependent spectrum function $Im A_C(\epsilon,t)$ where
$Im$ means taking the imaginary part of $A_C(\epsilon,t)$.
Notations $e$ and $\hbar$ denote, respectively, the electron
charge and Plank's constant. For the simplicity, we assume that
the tunneling rates $\Gamma_{L/R}$ are bias and energy
independent. $\Gamma_{L}$ and $\Gamma_{R}$ denote, respectively,
the tunneling rates from the left and right electrodes to the QD.
$f_{L(R)}(\epsilon)=1/(exp^{(\epsilon-\mu_{L(R)})/k_BT}+1)$ is the
Fermi distribution function of the left (right) electrode. The
chemical potential difference is related by the applied voltage
$\mu_L-\mu_R=eV_a$. From the results of Eqs. (3) and (4), the
time-dependent spectrum function plays a crucial role in the
determination of time-dependent tunneling current. The system
studied here considers that the step-like modulation is added into
the gate electrode, the expression of $A_C(\epsilon,t)$ is
obtained as

\begin{eqnarray}
A_C(\epsilon,t)& = & \frac{1}{\epsilon -(E_1-e V_g)+i\Gamma/2}\\
\nonumber & -&[ \frac{1}{\epsilon -(E_1-e V_g)+i\Gamma/2}
-\frac{1}{\epsilon -E_1+i\Gamma/2}]\\ \nonumber & \cdot
&exp^{i(\epsilon-(E_1-e V_g)+i\frac{\Gamma}{2}) t},
\end{eqnarray}
where $E_1$ is the ground state energy level of QD.
$\Gamma=\Gamma_L+\Gamma_R$. In the absence of gate voltage $V_g$
the $A_C(\epsilon)$ becomes time-independent retarded Green's
function $A_C(\epsilon)=1/(\epsilon-E_1+i\Gamma/2)$. The resonant
energy level $E_1$ is shifted to $E_1-e V_g$, that is
$A_C(\epsilon)=A_C(\epsilon)=1/(\epsilon-(E_1-e V_g) +i\Gamma/2)$,
when the system goes into steady state. In the transient process
the density of states of QD depends on time. Consequently, $J(t)$
displays time-dependent behavior in the transient process.

\section{Results and discussion}
Due to the complicate spectrum function, the transient current
lacks analytic form. To numerically calculate $J(t)$, we set the
Fermi energy level of electrodes and the ground state energy level
$E_F=50~meV$ and $E_1=90~meV$, respectively. Therefore, the energy
levels of QD are empty at zero temperature under zero bias. First
of all, we plot the time-dependent tunneling current through a
symmetric double-barrier tunneling structure with
$\Gamma_L=0.5~meV$ and $\Gamma_R=0.5~meV$ in response to a
step-like modulation for different applied voltages at zero
temperature in Fig. 1: solid line $(V_a=30~mV)$, dashed line
$(V_a=25~mV)$ and dotted line ($V_a=20~ mV$). The tunneling
current is zero at $t=0$ since the Fermi level of electrodes is
below the resonant state $E_1$. When the gate voltage is added
into the system, electrons from the left electrode are injected
into the new resonant energy level $E_1-eV_g$. Consequently, the
tunneling current jumps instantly and finally reaches the steady
state. Current also displays an interesting ringing behavior,
which was pointed out in Ref.[2]. The oscillations of the curve do
not maintain constant frequency in the transient process. In
addition, the frequency of oscillations is increased, when the
separation between the Fermi energy of left electrode and the
resonant level $E_1$ is decreased. Due to symmetry tunneling
rates, the behavior of $J(t)$ can be understood by the analysis of
$J^{in}_{L}(t)= J^{in}_{L,1}(t)+J^{in}_{L,2}(t)-J^{in}_R(t)$,where

\begin{eqnarray}
J^{in}_{L,1}(t)&=&\frac{-e}{\hbar} \Gamma_{L}
\int^{eV_a+E_F}_{eV_a}
\frac{d\epsilon}{\pi} \{ Im\frac{1}{\epsilon -(E_1-e V_g)+i\Gamma/2}\\
\nonumber & -& Im[ \frac{1}{\epsilon -(E_1-e V_g)+i\Gamma/2}
-\frac{1}{\epsilon -E_1+i\Gamma/2}]\\ \nonumber & \cdot &
cos((\epsilon-E_1+eV_g)t)exp^{-\frac{\Gamma}{2}t}\},
\end{eqnarray}
and
\begin{eqnarray}
J^{in}_{L,2}(t)&=&\frac{e}{\hbar} \Gamma_{L}
\int^{eV_a+E_F}_{eV_a} \frac{d\epsilon}{\pi} \{Real[
\frac{1}{\epsilon -(E_1-e V_g)+i\Gamma/2} \\ \nonumber &-&
\frac{1}{\epsilon -E_1+i\Gamma/2}]\cdot
sin((\epsilon-E_1+eV_g)t)exp^{-\frac{\Gamma}{2}t}\}.
\end{eqnarray}
We plot $J^{in}_{L,1}(t)$ (dashed line) and $J^{in}_{L,2}(t)$
(dotted line) in Fig. 2. $J^{in}_{L,1}(t)$ exhibits the behavior
of $(1-exp^{-\frac{\Gamma}{2}t})$, but $J^{in}_{L,2}(t)$ exhibits
the oscillation. Owing to the Fermi energy of right electrode away
from $E_1-eV_g$, we see small contribution from the $J^{in}_R(t)$.
From the results of Fig. 2, the oscillation of the solid curve
shown in Fig. 2 is consisted of two frequencies $\omega_1$ and
$\omega_2$, which are, respectively, given by $J^{in}_{L,2}(t)$
and $J^{in}_R(t)$.

Eq. (7) reveals that the frequency of oscillation depends on not
only $V_a$, but also $V_g$. To demonstrate it, we show $J(t)$ for
different applied gate voltages and $V_a=30~mV$ in Fig. 3: solid
line ($V_g=25~mV$), dashed line ($V_g=20~mV$), dotted line
($V_g=15~mV$) and dot-dashed line ($V_g=10~mV$). We see that the
frequency of oscillation can be tuned by the magnitude of applied
gate voltage. For $V_g=10~mV$, the Fermi energy level is just
aligned with $E_1-eV_g$, the oscillatory behavior of tunneling
current almost vanishes. Although the charge density of QD is not
be directly measured, we plot $N(t)$ in Fig. 4, where the curves
correspond to those of Fig. 3. It is very clear that QD is empty
before the gate voltage turns on. When the system reaches the
steady state, the value of $N(t)$ is fractional due to the result
of opened system. It is unexpect that $N(t)$ does not exhibit very
manifest oscillation feature as well as $J(t)$.

Because the width of tunneling barrier could not be precisely
controlled in the fabrication of SETs, we attempt to study the
tunneling current for the case of asymmetrical tunneling rates. We
show $J(t)$ for different tunneling rate ratio in Fig. 5: the
curves from the bottom to the top (for $t/t_0 < 0.5$) correspond
to $\Gamma_L=0.3, 0.4, 0.5,0.6$ and $0.7~meV$, respectively.
Meanwhile $\Gamma=\Gamma_L+\Gamma_R=1~meV$. Note that other
parameters used in Fig. 5 are the same as those of the solid line
of Fig. 1. It is worthy noting that the maximum steady current
occurs for the symmetry tunneling rates. This is consistent with
other theoretical reports$^{1}$. In addition, the steady current
of $\Gamma_L=0.7~meV$ ($\Gamma_L=0.6~meV$) is the same as that of
$\Gamma_L=0.3~meV$ ($\Gamma_L=0.4~meV$). This indicates that the
formula given by Eq. (1) maintains the asymmetry of $\Gamma_L$ and
$\Gamma_R$.

It is not avoided for high temperature SETs to study the
temperature effect on the transient current.  Figure 6 shows
$J(t)$ with symmetry tunneling rates for different temperatures at
$V_a=30~mV$ and $V_g=25~mV$. The ringing behavior of $J(t)$ is
swept out by temperatures. In addition, the magnitude of $J(t)$ is
suppressed with increasing temperature since the right electrode
provides current $J^{out}_{R}(t)$ into the QD, which is opposite
to $J^{in}_L(t)$. Consequently, the net current becomes small with
increasing temperature. We also show the charge density $N(t)$ in
Fig. 7: solid line ($k_BT=25~meV$) and dashed line
($k_BT=10~meV$). We see that $N(t)=N_L(t)+N_R(t)$ approaches the
same value for $k_BT=10~meV $ and $k_BT=25~meV$ in the steady
state. To understand this feature, we plot $N_L$ and $N_R$, which
are defined as the charge density provided from the left and right
electrodes, respectively. In the steady state (or a large time),
$N_L$ declines with increasing temperature, on the other hand
$N_R$ increases with increasing temperature. We see that they
compensate in each other.

Next, we study the transient current for the case of different
tunneling rate ratio $\Gamma_L/Gamma_R$ at room temperature
$k_BT=25~meV$. Comparing with the results at zero temperature, the
oscillation structure of $J(t)$ at room temperature is more
complicate due to the enhancement of $J^{in}_R(t)$. For
$\Gamma_L=0.7~meV$ and $\Gamma_L=0.3~meV$, the former exhibits the
exponential decay, the latter exhibits the exponential growth.
This indicates that the location of QD exists a considerable
effect on the transient transport properties of SETs. Even though
the above theoretical analysis is for SETs, we attempt to report
the experimental measurement of the transient current of Ge SHT.
For multivalley conduction band of germanium semiconductors, the
intervalley interaction effect can not be ignored. However, such
effect is not included in Eq. (1). The more realistic system
described by Eq. (1) is a Ge SHT, where valence band is a single
valley. In Figure 9 we show the transient current of Ge SHT for
two different applied voltages $V_a=20~mV$ and $V_a=30~mV$ at room
temperature and $V_g=-3~V$. The detailed fabrication process of Ge
SHTs and measurement technique of transient current will be
discussed in elsewhere$^{7}$. The solid lines are the experimental
curves. The dashed lines are the fitting curves. According to
fitting curves, the transient currents display exponential growth
feature. Based on previous theoretical analysis, the Ge SHT has
the characteristic of $\Gamma_L/\Gamma_R <~1$. Two fitting curves
also provide the bias-dependent tunneling times $70~sec$ and
$60~sec$ for the applied bias $V_a=20~mV$ and $V_a=30~mV$. This
very long tunneling time is arising from a fact of very high
barrier of $SiO_2$ which the Ge QD is embedded into. Besides, we
observe the small oscillation structures on the exponential growth
curves. The oscillation structures vanish for sufficient long
time. This indicates that this oscillation structures are not the
measurement error. Owing to the lack of information about the
detailed size and shape of QD, there are some difficulties to do
the detailed comparison between the theoretical calculation and
experimental measurement.

\section{Summary}
In this study the transient current of a SET is investigated by
using the formula derived by Jauho, Wingreen and Meir.$^{2}$ The
tunneling rate ratio exists a considerable influence on the
transient current of SETs. The tunneling current of a single Ge
SHT at room temperature displays the oscillation structures in the
transient process. This feature is attributed to time-dependent
density of states of QD in a switching on the gate voltage.

{\bf ACKNOWLEDGMENTS}

This work was supported by National Science Council of Republic of
China Contract Nos. NSC 94-2215-E-008-027 and NSC
94-2120-M-008-002.


\mbox{}


{\bf Figure Captions}

Fig. 1. Transient current through the symmetry barrier for
different applied voltages at zero temperature and $V_g=25~mV$:
solid line ($V_a=30~mV$), dashed line ($V_a=25~mV$) and dotted
line ($V_a=20~mV$), where $J_0=e~meV/\hbar$ and $t_0=\hbar/(meV)$.

Fig. 2. Transient current through the symmetry barrier for
$V_g=25~mV$, $V_a=30~mV$ and zero temperature. Solid line is
duplicated from that of Fig. 1. Dashed line and dotted line denote
$J^{in}_{L1}(t)$ and $J^{in}_{L2}$. Dash-dotted line represents
$J^{in}_{R}(t)$.

Fig. 3. Transient current through the symmetry barrier for
different applied gate voltages at zero temperature and
$V_a=30~mV$: solid line ($V_g=25~mV$), dashed line ($V_g=20~mV$),
dotted line ($V_g=15~mV$) and dash-dotted line ($V_g=10~mV$),
where $J_0/e=~meV/\hbar$ and $t_0=\hbar/(meV)$.

Fig. 4. Transient electron occupation number for different applied
gate voltages at zero temperature and $V_a=30~mV$. The curves are
one to one corresponding to those curves of Fig. 2.

Fig. 5. Transient current for different tunneling rate ratio at
zero temperature, $V_a=30~mV$ and $V_g=25`mV$: the curves from the
bottom to the top (for $t/t_0 <1$) correspond, respectively,
$\Gamma_L=0.3, 0.4, 0.5, 0.6$ and $0.7~meV$. Meanwhile,
$\Gamma=\Gamma_L+\Gamma_R=1~meV$. Here $J_0/e=~meV/\hbar$ and
$t_0=\hbar/(meV)$.

Fig. 6. Transient current through the symmetry barrier for
different temperatures at $V_a=30~mV$ and $V_g=25~mV$: solid line
($k_BT=0~meV$), dashed line ($k_BT=15~meV$), dotted line
($k_BT=20~meV$) and dash-dotted line ($k_BT=25~meV$), where
$J_0/e=~meV/\hbar$ and $t_0=\hbar/(meV)$.

Fig. 7. Transient electron occupation number for two different
temperatures: solid line ($k_BT=25~meV$) and dashed line
($k_BT=10~meV$).

Fig. 8. Transient current for different tunneling rate ratio  at
temperature $k_BT=25~meV$, $V_a=30~mV$ and $V_g=25`mV$.
$\Gamma=\Gamma_L+\Gamma_R=1~meV$. Here $J_0/e=~meV/\hbar$ and
$t_0=\hbar/(meV)$.

Fig. 9. Transient current of Ge single hole transistor for
different applied voltages at room temperature and $V_g=3~V$. The
curves from the bottom to the top correspond to $V_a=20~mV$ and
$30~mV$, respectively.


\end{document}